\documentclass[aps,twocolumn,prx,preprintnumbers,showpacs,amsmath,amssymb,superscriptaddress]{revtex4-1}
\usepackage[dvipdfm, colorlinks, citecolor=blue]{hyperref} 
\usepackage{graphicx}
\usepackage{dcolumn}
\usepackage{threeparttable}
\usepackage{multirow}
\usepackage{booktabs}
\usepackage{txfonts}
\usepackage{xcolor}
\usepackage{bm}
\usepackage{amssymb}
\usepackage{amsmath}
\usepackage{latexsym}
\usepackage{epsfig}
\usepackage{amsbsy}
\usepackage{array}
\usepackage{tabularx}
\usepackage{esvect} 
\usepackage{extarrows}

\newcommand{\rr}{\mathbf{r}}
\newcommand{\kk}{\mathbf{k}}
\newcommand{\ii}{\mathrm{i}}
\newcommand{\ee}{\mathrm{e}}
\newcommand{\GG}{\mathbf{G}}
\newcommand{\lG}{\mathbf{G}}
\newcommand{\qq}{\mathbf{q}}

\newcommand{\nk}{{n\mathbf{\kk}}}

\begin{document}

\title{Assessing model-dielectric-dependent hybrid functionals on the antiferromagnetic \\
transition-metal monoxides MnO, FeO, CoO, and NiO}

\author{Peitao Liu}
\email{peitao.liu@univie.ac.at}
\affiliation{University of Vienna, Faculty of Physics and Center for
Computational Materials Science, Sensengasse 8, A-1090 Vienna, Austria}

\author{Cesare Franchini}
\affiliation{University of Vienna, Faculty of Physics and Center for
Computational Materials Science, Sensengasse 8, A-1090 Vienna, Austria}
\affiliation{Dipartimento di Fisica e Astronomia, Universit\`{a} di Bologna, 40127 Bologna, Italy}

\author{Martijn Marsman}
\affiliation{University of Vienna, Faculty of Physics and Center for
Computational Materials Science, Sensengasse 8, A-1090 Vienna, Austria}

\author{Georg Kresse}
\affiliation{University of Vienna, Faculty of Physics and Center for
Computational Materials Science, Sensengasse 8, A-1090 Vienna, Austria}

\begin{abstract}
Recently,   two nonempirical hybrid functionals, dielectric-dependent range-separated hybrid functional based on the Coulomb-attenuating method (DD-RSH-CAM) and doubly screened hybrid functional (DSH), have been suggested  by  [\textcolor{blue}{Chen \emph{et al}, Phys. Rev. Mater. \textbf{2}, 073803 (2018)}] and [\textcolor{blue}{Cui \emph{et al}, J. Phys. Chem. Lett. \textbf{9}, 2338 (2018)}], respectively. These two hybrid functionals are both based on a common model dielectric function approach, but differ in the way how to non-empirically obtain the range-separation parameter. By retaining the full short-range Fock exchange and  a fraction of the long-range Fock exchange that equals the inverse of the dielectric constant, both DD-RSH-CAM and DSH turn out to perform very well in predicting the band gaps for a large variety of  semiconductors and insulators. Here, we  assess how these two hybrid functionals perform on challenging antiferromagnetic transition-metal monoxides MnO, FeO, CoO, and NiO by comparing them to other conventional hybrid functionals and the $GW$ method. We find that single-shot DD0-RSH-CAM and DSH0 improve the band gaps towards experiments as compared to conventional hybrid functionals. The magnetic moments are slightly increased, but the predicted dielectric constants are decreased. The valence band density of states (DOS) predicted by DD0-RSH-CAM and DSH0 are as satisfactory as HSE03 in comparison to experimental spectra, however, the conduction band DOS are shifted to higher energies by about 2 eV compared to HSE03. Self-consistent DD-RSH-CAM and DSH deteriorate the results with a significant overestimation of band gaps.
\end{abstract}

\maketitle

\section{Introduction}\label{sec:intro}

Kohn-Sham (KS) density functional theory (DFT)~\cite{PhysRev.136.B864, PhysRev.140.A1133} in the local density
approximation (LDA) and semilocal generalized gradient approximation (GGA)~\cite{PhysRevB.46.6671}
continues to be a very powerful and widely used tool for the quantitative prediction of
ground-state properties of materials in solid-state physics as well as in chemistry owing to
its reasonably good accuracy and relatively low computational cost.
Nevertheless, excited states, which go beyond the abilities of
standard KS-DFT, are usually poorly described. For instance, KS-DFT
in the LDA and GGA  always underestimates the band gaps compared to the experimental values~\cite{PhysRevLett.51.1888}.
This is due to the so-called many-electron ``self-interaction error (SIE)"  inherent
to LDA or GGA functionals~\cite{doi:10.1063/1.2403848,RevModPhys.86.253}.
The generally established method to calculate the quasiparticle energies and band gaps is the state-of-the-art
$GW$ method~\cite{PhysRev.139.A796}. Nevertheless, $GW$ calculations
are computationally much more  demanding than DFT-based methods.
Although the cubic-scaling $GW$ method has been recently developed~\cite{PhysRevB.94.165109},
the large prefactor compared to DFT makes it still difficult to apply to very large extended systems.
In addition, the high cost required in calculating the forces in
the random phase approximation (RPA)~\cite{PhysRevLett.118.106403} limits the application of the $GW$ method
for structure relaxations.

The other widely used methods that can cure the band gap problem are hybrid functionals,
which are constructed by admixing a fraction of exact nonlocal Fock exchange to
a (semi)local exchange-correlation (XC) potential, e.g., the Perdew-Burke-Ernzerhof (PBE) functional~\cite{PhysRevLett.77.3865},
reducing the SIE.
Apart from the band gap, hybrid functionals can also give good descriptions for total energies and forces.
The PBE0 hybrid functional~\cite{doi:10.1063/1.472933,doi:10.1063/1.478522},
which includes one quarter of the Fock exchange, reproduces the homogeneous electron gas limit
and significantly outperforms the B3LYP hybrid functional~\cite{doi:10.1063/1.464304} in solids, in particular, in
systems with itinerant character such as metals and small-gap semiconductors~\cite{doi:10.1063/1.2747249}.
However, calculating the long-range (LR) exchange interactions in PBE0 is
computationally demanding, and particularly difficult for metals
where a dense $k$-point sampling is required, resulting in very slow convergence. To address this issue,
Heyd, Scuseria, and Ernzerhof  (HSE)~\cite{doi:10.1063/1.1564060} proposed to replace the LR Fock exchange
by the corresponding density functional counterpart, e.g., PBE exchange.
The proposed hybrid functional with a range-separation parameter $\mu=0.2~{\AA}^{-1}$ was
referred to as HSE03~\cite{doi:10.1063/1.1564060}. Afterwards, the HSE06 hybrid functional with $\mu=0.3~{\AA}^{-1}$
was suggested~\cite{doi:10.1063/1.2404663}.
Nevertheless,  the HSE06 predicted band gaps are not satisfactory for weakly screening large band gap materials.
To improve the band gaps, a modified HSE with a short-range (SR) Fock exchange fraction
$\alpha=0.6$ and $\mu=0.5~{\AA}^{-1}$ was proposed in Ref.~\cite{RevModPhys.86.253}.
We denote this modified HSE as MHSE.

So far $\alpha$ and $\mu$ have been taken as empirical parameters, which depend on the specific functional used
and the system studied.  They can be adjusted to better ``fit"  the  experimental results~\cite{PhysRevB.86.235117, Franchini_2014}.
The dependence of the results on the choice of these parameters limits the predictive capability of these functionals.
To overcome this problem, a full-range dielectric-dependent hybrid (DDH) functional
has been suggested. It has the same form as PBE0, but the fraction of the Fock exchange is determined
by the inverse of the static dielectric constant $\epsilon^{-1}_\infty$,
based on the connection between hybrid functionals and the static Coulomb hole plus screened exchange (COHSEX)
approximation~\cite{doi:10.1002/pssb.201046195, PhysRevB.83.035119,doi:10.1063/1.4722993}.
To further improve the band gaps of typical $sp$ insulating materials, self-consistent DDH (sc-DDH) has been
proposed~\cite{PhysRevB.89.195112}, which determines the Fock exchange fraction in a self-consistent manner.
Sc-DDH has been widely used to describe band gaps of oxide semiconductors~\cite{PhysRevB.91.155201,0953-8984-29-45-454004,0953-8984-30-4-044003},
defects~\cite{doi:10.1063/1.4931805,doi:10.1021/acs.jpcc.6b02707}, band alignments of semiconductors~\cite{PhysRevB.95.075302},
and interfaces~\cite{doi:10.1021/acs.jctc.7b00853,PhysRevMaterials.3.073803}.

Although DDH performs well for wide-gap insulators, it shows large errors
for systems with narrow band gaps~\cite{0953-8984-29-45-454004,PhysRevB.93.235106}
due to the neglect of the range dependency in the screened exchange potential.  To address this issue,
Skone \emph{et al.}~\cite{PhysRevB.93.235106} proposed a range-separated DDH (RS-DDH),
where the range-separation parameter $\mu$ is determined
by fitting the calculated dielectric functions from first-principles with
model dielectric functions~\cite{PhysRevB.47.9892,SHIMAZAKI200891,RevModPhys.86.253}.
Nevertheless, in RS-DDH, the fraction of the LR Fock exchange is fixed to $\beta=\epsilon^{-1}_\infty$,
whereas the SR fraction is empirically set to $\alpha=1/4$ as in PBE0.
To eliminate this empirical setting of the SR Fock fraction,  very recently,   two nonempirical hybrid functionals,
dielectric-dependent range-separated hybrid functional based on the Coulomb-attenuating
method (DD-RSH-CAM)~\cite{PhysRevMaterials.2.073803}
and doubly screened hybrid functional (DSH)~\cite{doi:10.1021/acs.jpclett.8b00919}
have been proposed independently.
Despite of different motivations,
these two hybrid functionals have essentially the same expression with a common  model dielectric function,
but differ in the way how to non-empirically obtain the range-separation parameter $\mu$.
By keeping the full SR Fock exchange and including a $\beta=\epsilon^{-1}_\infty$ fraction of the LR Fock exchange,
both, self-consistent DD-RSH-CAM and DSH,  turn out to perform very well in predicting the band gaps
for a large variety of semiconductors and insulators with narrow, intermediate, or
wide gaps~\cite{PhysRevMaterials.2.073803,doi:10.1021/acs.jpclett.8b00919}.
In addition,  it is found that the simplified single-shot DD0-RSH-CAM and DSH0 with $\epsilon_\infty$
computed from the PBE, one-electron energies and orbitals at the level of RPA
almost perform equally well as their self-consistent
counterparts~\cite{PhysRevMaterials.2.073803,doi:10.1021/acs.jpclett.8b00919}.

In this paper, we identify the connections of above-mentioned hybrid functionals
within the generalized Kohn-Sham (gKS) formalism~\cite{PhysRevB.53.3764}
and assess how these two recently proposed promising hybrid functionals, DD-RSH-CAM and DSH,
in single-shot and self-consistent versions, perform on
more challenging antiferromagnetic transition-metal (TM) monoxides MnO, FeO, CoO, and NiO
in terms of dielectric constants, band gaps, and magnetic moments as well as density of states (DOS),
by comparing them to other conventional hybrid functionals such as PBE0, HSE03, HSE06, and MHSE, as well as the $GW$ method.
The reasons that we have chosen these four TM monoxides as our systems of study
are threefold. First, these four compounds have been extensively studied
in experiments~\cite{PLENDL1969109,doi:10.1063/1.1714508,PhysRevB.44.1530,
PhysRevB.44.6090,PhysRevLett.53.2339,PhysRevB.77.165127,MnO1968,BOWEN1975355,PhysRevB.2.2182,jkps.50.632,
PhysRevB.27.6964,PhysRev.110.1333,PhysRevB.1.2243,Alperin1962,0022-3719-12-2-021,PhysRevB.64.052102}
so that there are many available experimental data to compare to.
Second, these four compounds are considered to be prototypical strongly
correlated electron systems and thus have been taken to be testbed materials for
many new theoretical methods, e.g.,  ranging from LSDA~\cite{PhysRevB.30.4734,0953-8984-11-7-002},
LSDA+$U$~\cite{HUGEL1996457, PhysRevB.49.10864, PhysRevB.55.12822, PhysRevB.71.035105, PhysRevB.74.155108},
HSE~\cite{PhysRevB.74.155108,PhysRevB.72.045132, PhysRevB.75.195128, 0953-8984-20-6-064201},
$GW$~\cite{PhysRevLett.74.3221, PhysRevLett.93.126406, PhysRevB.79.235114,PhysRevB.82.045108,PhysRevB.86.235122}
to DFT+DMFT (dynamical mean-field theory)~\cite{PhysRevB.74.195114, PhysRevLett.109.186401,PhysRevB.96.045111}.
Third, to our knowledge from literature, using a single method,  it seems to be impossible to describe well the band
gaps for all four compounds simultaneously. For instance, though HSE03 describes the band gaps for FeO and NiO
in good agreement with experiments, it significantly underestimates the band gap of MnO by  1.1 eV and overestimates
the band gap of CoO by 0.9 eV~\cite{0953-8984-20-6-064201}. Single-shot $G_0W_0$ on top of HSE03 improves the band
gap of MnO, but it yields an even larger band gap for CoO~\cite{PhysRevB.79.235114}.
In addition, the band gap of NiO is increased by $G_0W_0$@HSE03, now deviating from experiment~\cite{PhysRevB.79.235114}.

Compared to conventional hybrid functionals, we find that single-shot DD0-RSH-CAM and DSH0
give an excellent description of the band gaps for all four compounds with the
smallest mean absolute percentage error (MAPE) with respect to experimental gaps.
In addition, the magnetic moments are found to increase very slightly ($<0.1~\mu$B/atom) compared to conventional hybrid functionals.
On the other hand, DD0-RSH-CAM and DSH0 decrease the dielectric constants, deviating from experiments.
Moreover, it is found that DD0-RSH-CAM and DSH0, similar to HSE03, yield good valence band DOS
in comparison to the experimental spectra, but the predicted conduction band DOS are shifted
to higher energies than for HSE03. Furthermore, in contrast to what is observed in
Refs.~\cite{PhysRevMaterials.2.073803,doi:10.1021/acs.jpclett.8b00919},
self-consistent DD-RSH-CAM and DSH deteriorate the results for all these four compounds
with a substantial overestimation of band gaps.

The paper is organized as follows. In Sec.~\ref{sec:method},
we will make a short summary of all above-mentioned hybrid functionals
and identify the connections among them. Particular emphasis
is devoted to the model dielectric function that DD-RSH-CAM and DSH have in common
and how the range-separation parameter $\mu$ is determined in each of these two hybrid functionals.
Technical details and computational setups will be provided in
Sec.~\ref{sec:details}. The results will be presented and discussed in
Sec.~\ref{sec:results} and summarized in Sec.~\ref{sec:conlcusions}.

\section{Theoretical background}\label{sec:method}
Within the gKS formalism~\cite{PhysRevB.53.3764}, the total potential $V_{\rm gKS}(\rr,\rr')$ reads:
\begin{eqnarray}
V_{\rm gKS}(\rr,\rr') = V_{\rm H}(\rr) + V_{\rm ext}(\rr) + V_{\rm xc}(\rr,\rr'),
\end{eqnarray}
where $V_{\rm H}$, $V_{\rm ext}$, and $V_{\rm xc}$ are the Hartree,  external,  and exchange-correlation potential, respectively.
The nonlocal $V_{\rm xc}$ is made up by the full density functional  correlation potential $V^{\rm PBE}_{\rm c}$ (here always PBE), and
admixing a certain amount of the nonlocal Fock exchange $V^{\rm Fock}_{\rm x}$ to the semilocal PBE exchange $V^{\rm PBE}_{\rm x}$~\cite{PhysRevB.89.195112}
\small
\begin{equation}\label{eq:gKS}
\begin{split}
& V_{\rm xc}(\rr,\rr')  = \alpha V^{\rm Fock, SR}_{\rm x}(\rr, \rr' ; \mu) + (1-\alpha) V^{\rm PBE, SR}_{\rm x}(\rr; \mu)\delta(\rr-\rr') \\
& + \beta V^{\rm Fock, LR}_{\rm x}(\rr, \rr' ; \mu) + (1-\beta) V^{\rm PBE, LR}_{\rm x}(\rr; \mu)\delta(\rr-\rr')+V^{\rm PBE}_{\rm c}(\rr)\delta(\rr-\rr').
\end{split}
\end{equation}
\normalsize
Here, $\alpha$ and $\beta$ denote the fraction of the SR and LR Fock exchange, respectively.
The SR and LR exchange potentials are determined by partitioning the Coulomb potential~\cite{doi:10.1063/1.1564060} according to
\small
\begin{eqnarray}\label{eq:range-separation}
    \frac{1}{ |\rr-\rr'| }
=\underbrace{\frac{\text{erfc}(\mu |\rr-\rr'|)}{ |\rr-\rr'| }}_{\text{\text{SR}}}+\underbrace{\frac{\text{erf}(\mu |\rr-\rr'|)}{ |\rr-\rr'| }}_{\text{\text{LR}}},
\end{eqnarray}
\normalsize
where $\mu$ is the range-separation parameter. It is related to a characteristic distance $2/\mu$
beyond which the SR interactions become negligible.
With Eq.~(\ref{eq:range-separation}), $V^{\rm Fock, SR}_{\rm x}(\rr, \rr' ; \mu)$ and $V^{\rm Fock, LR}_{\rm x}(\rr, \rr' ; \mu)$ are expressed as:
\small
\begin{eqnarray}
V^{\rm Fock, SR}_{\rm x}(\rr, \rr' ; \mu) &=&  -e^2 \sum_{\nk}  w_\kk f_\nk   \psi^*_\nk(\rr') \psi_\nk(\rr)\frac{ \text{erfc}(\mu |\rr-\rr'|) }{|\rr-\rr'|}, \\
V^{\rm Fock, LR}_{\rm x}(\rr, \rr' ; \mu) &=&  -e^2 \sum_{\nk}  w_\kk f_\nk   \psi^*_\nk(\rr') \psi_\nk(\rr)\frac{ \text{erf}(\mu |\rr-\rr'|) }{|\rr-\rr'|}.
\end{eqnarray}
\normalsize
Here, $\psi_\nk(\rr)$ are one-electron Bloch states of the system and $f_\nk$ are their corresponding occupation numbers.
The sum over $\kk$ is performed over all $\kk$ points used to sample the Brillouin zone (BZ)
and the sum over $n$ is performed over all bands at these $\kk$ points with corresponding weights $w_\kk$.

Comparing the XC potential
of the PBE0 hybrid functional~\cite{doi:10.1063/1.472933,doi:10.1063/1.478522}
\small
\begin{equation}\label{eq:PBE0}
\begin{split}
V^{\rm PBE0}_{\rm xc}(\rr,\rr')  = \frac{1}{4} V^{\rm Fock}_{\rm x}(\rr, \rr') + \frac{3}{4} V^{\rm PBE}_{\rm x}(\rr)\delta(\rr-\rr')
+V^{\rm PBE}_{\rm c}(\rr)\delta(\rr-\rr')
\end{split}
\end{equation}
\normalsize
to Eq.~(\ref{eq:gKS}), one can see that PBE0
corresponds to $\alpha=\beta=1/4$.
DDH has the same form as PBE0, but the mixing parameters are determined
by the inverse of the static dielectric constant, i.e., $\alpha=\beta=\epsilon^{-1}_\infty$.
RS-DDH~\cite{PhysRevB.93.235106} can be obtained by setting $\alpha=1/4$ and $\beta=\epsilon^{-1}_\infty$.
On the other hand, the HSE03 hybrid functional~\cite{doi:10.1063/1.1564060} can be recovered for
 $\alpha=1/4$, $\beta=0$ and $\mu=0.2~{\AA}^{-1}$.
Similarly, the HSE06 hybrid functional~\cite{doi:10.1063/1.2404663} is equivalent to
$\alpha=1/4$, $\beta=0$ and $\mu=0.3~{\AA}^{-1}$
and the MHSE proposed in Ref.~\cite{RevModPhys.86.253} corresponds to $\alpha=0.6$, $\beta=0$ and $\mu=0.5~{\AA}^{-1}$.

Furthermore, the two recently proposed model-dielectric-dependent hybrid functionals,
DD-RSH-CAM~\cite{PhysRevMaterials.2.073803} and DSH~\cite{doi:10.1021/acs.jpclett.8b00919}, are  equivalent to
$\alpha=1$ and $\beta=\epsilon^{-1}_\infty$. The explicit form of the XC potential is expressed as:
\small
\begin{equation}\label{eq:modified_hybridExc_V}
\begin{split}
V_{\text{xc}}(\rr, \rr')
&= V_{\text x}^{\text{Fock ,SR}}(\rr, \rr'; \mu) +\epsilon^{-1}_\infty V_{\text x}^{\text{Fock,LR}}(\rr,\rr'; \mu) \\
&  + (1-\epsilon^{-1}_\infty) V_{\text x}^{\text{PBE,LR}}(\rr; \mu)\delta(\rr-\rr')+ V_{\text c}^{\text{PBE}}(\rr)\delta(\rr-\rr'), \\
&=\Big[1-(1-\epsilon^{-1}_\infty) {\rm erf}(\mu|\rr-\rr'|)\Big] V_{\text x}^{\text{Fock}}(\rr, \rr') \\
& + (1-\epsilon^{-1}_\infty) V_{\text x}^{\text{PBE,LR}}(\rr; \mu)\delta(\rr-\rr') + V_{\text c}^{\text{PBE}}(\rr)\delta(\rr-\rr').
\end{split}
\end{equation}
\normalsize
$V^{\rm Fock}_{\rm x}$ is the full-range Fock exchange,
which is given by
\small
\begin{equation}
\begin{split}
V^{\rm Fock}_{\rm x} (\rr, \rr')= -e^2 \sum_{\nk} w_\kk f_\nk  \frac{  \psi^*_\nk(\rr') \psi_\nk(\rr)}{|\rr-\rr'|}.
\end{split}
\end{equation}
\normalsize
The representation of $\small V(\rr,\rr')=\Big[1-(1-\epsilon^{-1}_\infty) {\rm erf}(\mu|\rr-\rr'|)\Big] V_{\text x}^{\text{Fock}}(\rr, \rr')\normalsize $  in Eq.~(\ref{eq:modified_hybridExc_V})
 in reciprocal space is given by
\small
\begin{equation}\label{eq:MFOCK_V}
\begin{split}
& V_\qq(\lG,\lG')=\frac{1}{\Omega} \iint d\rr d\rr'  \ee^{-\ii (\qq +\GG) \cdot \rr}V(\rr,\rr') \ee^{\ii (\qq +\GG') \cdot \rr'} =\\
&  -\frac{4\pi e^2}{\Omega} \sum_\nk w_\kk f_\nk
 \sum_{\GG''}   \frac{C^*_\nk(\GG'-\GG'')C_\nk(\GG-\GG'')}{|\qq-\kk+\GG''|^2} \cdot \epsilon^{-1}(|\qq-\kk+\GG''|),
\end{split}
\end{equation}
\normalsize
where $\Omega$ is the volume of the system and $C_\nk(\GG)$ are the plane-wave expansion coefficients
of Bloch states $\psi_\nk(\rr)=1/\sqrt{\Omega} \sum_\lG C_\nk(\GG) \ee ^{\ii (\kk+\lG)\cdot \rr}$.
The ``model dielectric function" $\epsilon(|\lG|)$ here is defined by
\begin{equation}\label{eq:model_die}
\begin{split}
\epsilon^{-1}(|\lG|)=1-(1-\epsilon^{-1}_{\infty}) \ee^{-|\lG|^2/4\mu^2}.
\end{split}
\end{equation}
This is in contrast to the model dielectric function  $\epsilon^{-1}(|\lG|)=\frac{1}{4} [1-\ee^{-|\GG|^2/4\mu^2}]$ used in HSE~\cite{doi:10.1063/1.2187006}.
This model dielectric function Eq.~(\ref{eq:model_die}) has also been used to approximate
the static screened interaction $W$ in the Bethe-Salpeter equation (BSE)
in simulating optical spectra~\cite{Bokdam2016, PhysRevMaterials.2.075003}.

\begin{table}
\caption {Comparison of different hybrid functionals
in terms of the SR Fock exchange fraction $\alpha$,  the LR Fock exchange fraction $\beta$,
the range-separation parameter $\mu$ (in $\AA^{-1}$) and the model dielectric function $\epsilon^{-1}(|\GG|)$.
}
\begin{ruledtabular}
\begin{tabular}{lllll}
        &   $\alpha$    &   $\beta$       &  $\mu$  &  $\epsilon^{-1}(|\GG|)$   \\
        \hline
 PBE0   &   $1/4$              &   $1/4$                 &  0          &  $1/4$  \\
 HSE03  &   $1/4$              &   0                &  0.3        &  $\frac{1}{4}[1-\ee^{-|\GG|^2/4\mu^2}]$ \\
 HSE06  &   $1/4$              &   0                 &  0.2        &  $\frac{1}{4} [1-\ee^{-|\GG|^2/4\mu^2}]$ \\
 MHSE  &   $0.6$              &   0                 &  0.5        &  $0.6 [1-\ee^{-|\GG|^2/4\mu^2}]$ \\
 DDH    & $\epsilon^{-1}_{\infty}$ & $\epsilon^{-1}_{\infty}$  & 0 &  $\epsilon^{-1}_{\infty}$ \\
 RS-DDH & $1/4$    & $\epsilon^{-1}_{\infty}$  & from fit  & $\frac{1}{4}-(\frac{1}{4}-\epsilon^{-1}_{\infty})\, \ee^{-|\GG|^2/4\mu^2}$\\
 DD-RSH-CAM &   1              &   $\epsilon^{-1}_{\infty}$      &  from fit           &   $1-(1-\epsilon^{-1}_{\infty})\, \ee^{-|\GG|^2/4\mu^2}$ \\
 DSH    &   1              &   $\epsilon^{-1}_{\infty}$      &  from (\ref{eq:mu_for_DSH})   & $1-(1-\epsilon^{-1}_{\infty})\, \ee^{-|\GG|^2/4\mu^2}$
\end{tabular}
\end{ruledtabular}
\label{Table: HSE_MDDH_compare}
\end{table}

\begin{figure}
\begin{center}
\includegraphics[width=0.48\textwidth,clip]{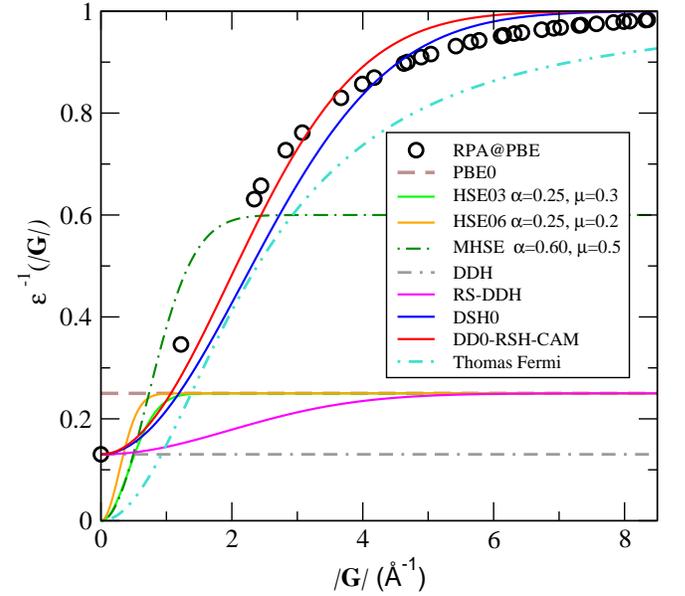}
\end{center}
 \caption{ (Color online) Model dielectric functions $\epsilon^{-1}(|\lG|)$ for different hybrid functionals.
The circles indicate the computed dielectric functions of MnO from first-principles at the level of RPA@PBE.
The parameters $\beta$ and $\mu$ used in single-shot DD0-RSH-CAM and DSH0 are given in Table~\ref{Table: beta_mu}.
RS-DDH uses the same $\mu$ as in DD0-RSH-CAM.
The Thomas-Fermi screening is given by $\epsilon^{-1}(|\lG|)=b |\lG|^2/(b|\lG|^2+q_{\rm TF}^2)$~\cite{PhysRevB.41.7868}.
Here, an empirical parameter  $b= 1.563$ is used~\cite{PhysRevB.47.9892}.
}
\label{fig:HSE_eps_G}
\end{figure}

For DD-RSH-CAM, $\mu$ is obtained in a similar way to DD-RSH
through least-squares fitting to the inverse of the dielectric function
in the long-wavelength limit $\epsilon^{-1}_{\lG,\lG}(\qq\rightarrow0, \omega=0)$
using the model dielectric function Eq.~(\ref{eq:model_die})~\cite{PhysRevMaterials.2.073803}.
However, for DSH, $\mu$ is approximated by~\cite{doi:10.1021/acs.jpclett.8b00919}
\begin{eqnarray}\label{eq:mu_for_DSH}
\mu = \frac{2q^2_{\rm TF}}{3b(1-\epsilon^{-1}_\infty)}.
\end{eqnarray}
Here, $q^2_{\rm TF}=4 \Big(\frac{3\rho}{\pi}\Big)^{1/3}$ denotes the Thomas-Fermi screening parameter,
where $\rho$ is the valence electron density of the system.
The empirical parameter $b=1.563$ is suggested to better capture the dielectric function
of typical semiconductors~\cite{PhysRevB.47.9892, doi:10.1021/acs.jpclett.8b00919}.
From Eq.~(\ref{eq:mu_for_DSH}) one can see that $\mu$ for DSH increases as $\epsilon_\infty$ decreases.

Table~\ref{Table: HSE_MDDH_compare} summarizes a comparison between different hybrid functionals
in terms of $\alpha$, $\beta$, $\mu$, and the model dielectric function $\epsilon^{-1}(|\GG|)$ used.
Taking MnO as an example, Fig.~\ref{fig:HSE_eps_G} furthermore shows the model dielectric functions
of different hybrid functionals, along with the Thomas-Fermi screening. Clearly, in describing the screening of MnO,
the conventional hybrid functionals such as PBE0, HSE03 and HSE06 as well as MHSE, DDH,  and RS-DDH
are all unsatisfactory, whereas single-shot DD0-RSH-CAM and DSH0 are almost equally good, since
they reproduce the momentum-dependent screening from first-principles.
Thus, it is not surprising that DD0-RSH-CAM and DSH0 outperform the conventional hybrid functionals
in the description of the band gaps for various semiconductors and insulators spanning a wide range
of band gaps~\cite{PhysRevMaterials.2.073803,doi:10.1021/acs.jpclett.8b00919}.

\section{Computational details}\label{sec:details}

The DD-RSH-CAM and DSH hybrid functionals within the projector augmented
wave (PAW) formalism~\cite{PhysRevB.50.17953} have been implemented in the Vienna
\emph{Ab initio} Simulation Package (VASP)~\cite{PhysRevB.47.558, PhysRevB.54.11169}
with little effort based on existing routines of the screened hybrid functionals~\cite{doi:10.1063/1.2187006}.
We note that the implementation has been available since VASP.5.2 in 2009, although the gradient contribution
is only properly implemented in VASP.6 and has been made available by the authors of Ref.~\cite{doi:10.1021/acs.jpclett.8b00919}.
For the (semi)local part of the exchange and correlation, the PBE XC functional was used.
The Mn\_sv, Fe\_sv, Co\_sv, Ni\_pv and standard O PAW potentials were used for PBE and hybrid functional calculations. Specifically,
the oxygen 2$s$ and 2$p$ electrons as well as the 3$s$, 3$p$, 3$d$, and 4$s$ electrons of the Mn, Fe, and Co atoms
are treated as valence states. For the Ni atom, 3$p$, 3$d$, and 4$s$  electrons
are treated as valence states.
The plane-wave cutoff for the orbitals was chosen to be the maximum
of all elements in the considered material. 8$\times$8$\times$8 $\Gamma$-centered
$k$-point grids were used to sample the BZ. Spin polarization was considered, but spin-orbit coupling was not included.
For all the four TM monoxides, the experimental rock-salt
crystal structures in the ground-state type-II antiferromagnetic (AFII) ordering
with lattice constants of 4.445~$\AA$, 4.334~$\AA$,  4.254~$\AA$,
and 4.171~$\AA$ for MnO, FeO, CoO, and NiO, respectively,
were used~\cite{MnO_expt_str, FeO_expt_str, CoO_expt_str,NiO_expt_str}.
Note that the experimentally observed small distortions~\cite{PhysRevB.27.6964,PhysRevB.64.052102}
were neglected in our calculations.
However, it is important to note that due to the degenerate high-spin ground-state configurations of
${\rm Fe}^{2+}(t^{\uparrow 3}_{\rm 2g}e^{\uparrow 2}_{\rm g}t^{\downarrow 1}_{\rm 2g})$
and ${\rm Co}^{2+}(t^{\uparrow 3}_{\rm 2g}e^{\uparrow 2}_{\rm g}t^{\downarrow 2}_{\rm 2g})$
in a octahedral crystal field, any band theory would fail to open the band gaps for FeO and CoO.
To address this issue,  we have manually broken the $t_{\rm 2g}$ degeneracies by slightly distorting the lattice.

To determine  the LR Fock exchange fraction $\beta$ in DD-RSH-CAM and DSH,
calculations of the static dielectric constants $\epsilon_\infty$ are needed.
For single-shot DD0-RSH-CAM and DSH0, one computes $\epsilon_\infty$ by RPA
using the PBE one-electron energies and orbitals~\cite{PhysRevB.73.045112}. RPA@PBE
describes screening properties reasonably well for semiconductors~\cite{PhysRevB.78.121201}
due to fortuitous cancellation of errors originating from the underestimated PBE
band gap and the absence of electron-hole interactions.
For self-consistent DD-RSH-CAM,  we followed the strategy given in Ref.~\cite{PhysRevMaterials.2.073803},
i.e., by including a nonlocal XC kernel $f_{\rm xc}$. This is necessary because,
if the band gap is close to experiment, e.g., for hybrid functionals, then the RPA will significantly underestimate the dielectric constant.
For $f_{\rm xc}$ in DD-RSH-CAM, the bootstrap approximation of Sharma \emph{et al.}~\cite{PhysRevLett.107.186401} $f^{\rm boot}_{\rm xc}$ is employed.
For DD-RSH-CAM, the parameter $\mu$  is always determined by fitting $\epsilon^{-1}_{\lG,\lG}(\qq\rightarrow0, \omega=0)$
from first-principles [RPA@PBE for DD0-RSH-CAM and (RPA+$f^{\rm boot}_{\rm xc}$)@DD-RSH-CAM for self-consistent DD-RSH-CAM]
through the model dielectric function Eq.~(\ref{eq:model_die}).

For self-consistent DSH, we adopted the finite field
approach~\cite{PhysRevB.63.155107,PhysRevLett.89.117602} as in Ref.~\cite{doi:10.1021/acs.jpclett.8b00919},
except for FeO, for which we have used the linear response theory method as adopted in the self-consistent DD-RSH-CAM calculations
due to critical difficulties in converging $\epsilon_\infty$ for FeO by the finite field approach.
The finite field approach calculates the static dielectric tensor from the change in the polarization due to small but finite electric fields
and essentially includes all self-consistent  local field effects from self-consistent changes of the Hartree (RPA) as well XC potential~\cite{PhysRevB.78.121201}.
This approach yields within the approximations made by the functional essentially the {\em exact} $\epsilon_\infty$ and is
hence fundamentally more accurate than the approach taken in DD-RSH-CAM, where the approximate
bootstrap kernel  $f^{\rm boot}_{\rm xc}$ is used.
On the other hand, for DSH $\mu$  is simply approximated through Eq.~(\ref{eq:mu_for_DSH}) which is potentially more approximate. The parameter $\mu$
is updated whenever $\epsilon_\infty$ changes.
Self-consistency for DD-RSH-CAM and DSH is achieved when the change in $\epsilon_\infty$
in two sequential iterations is less than 0.01.

For comparison, single-shot $G_0W_0$ and eigenvalue self-consistent ev$GW_0$~\cite{PhysRevB.75.235102}
on top of PBE were also performed.  For the $GW$ calculations, The $GW$ PAW potentials, i.e.,
Mn\_sv\_GW, Fe\_sv\_GW, Co\_sv\_GW, Ni\_sv\_GW and O\_GW, were adopted.
The energy cutoff for the response function was set to be half of the energy cutoff for the orbtials,
and the total number of orbitals was chosen to be equal to the number of plane-waves.

\section{Results and discussions}\label{sec:results}

\begin{table}
\caption {LR Fock exchange fraction $\beta$ and range-separation parameter $\mu$ (in $\AA^{-1}$)
used in single-shot DD0-RSH-CAM and DSH0, and converged self-consistent DD-RSH-CAM and DSH calculations. }
\begin{ruledtabular}
\begin{tabular}{clcccc}
  &       &   MnO  &   FeO       &  CoO   &   NiO  \\
         \hline
\multirow{4}{0.1cm}{$\beta$}
& DD0-RSH-CAM   & 0.13 & 0.00 & 0.00 & 0.06 \\
& DSH0 & 0.13 & 0.00 & 0.00 & 0.06 \\
& DD-RSH-CAM & 0.28 & 0.28 & 0.28 & 0.28 \\
& DSH & 0.26 & 0.27 & 0.26 & 0.25 \\
\hline
\multirow{4}{0.1cm}{$\mu$}
& DD0-RSH-CAM   & 1.39 & 1.42 & 1.44 & 1.43 \\
& DSH0 & 1.55 &  1.48 & 1.50 & 1.55 \\
& DD-RSH-CAM & 1.46 & 1.55 & 1.56 & 1.56 \\
& DSH & 1.67 & 1.73 & 1.62 & 1.74
\end{tabular}
\end{ruledtabular}
\label{Table: beta_mu}
\end{table}

\begin{figure}
\begin{center}
\includegraphics[width=0.48\textwidth,clip]{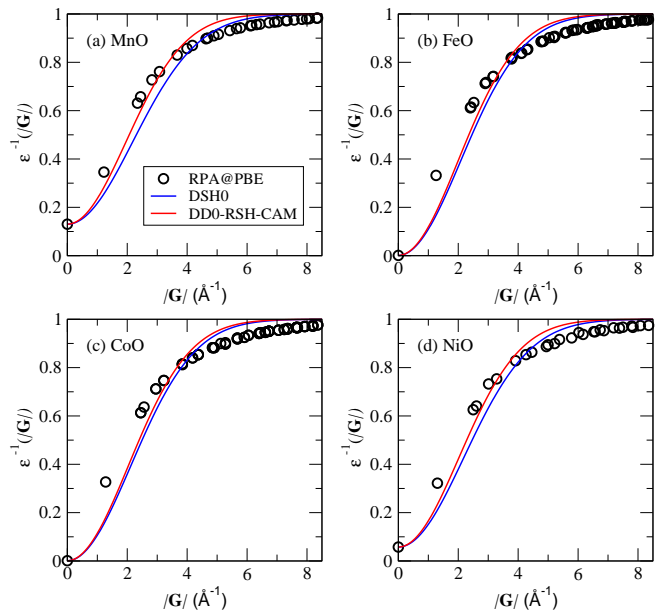}
\end{center}
 \caption{(Color online) Inverse dielectric function $\epsilon^{-1}_{\lG,\lG}(\qq\rightarrow0, \omega=0)$ calculated from RPA@PBE (circles)
 compared to the model dielectric function in DD0-RSH-CAM (red lines) and DSH0 (blue lines) for (a) MnO, (b) FeO, (c) CoO, and (d) NiO.
}
\label{fig:eps_G_all}
\end{figure}

\begin{table*}
\caption {
Indirect ($E^i_{\rm g}$)  and direct ($E^d_{\rm g}$) band gaps (in eV)
calculated from different theoretical approaches as well as available experimental gaps.
The mean absolute percentage error (MAPE) with respect to experiment is also given.
Note that in calculating the MAPE, we have used the deviations between theoretical indirect gaps
and averaged experimental PES+BIS and XAS+XES fundamental gaps, except for FeO,
for which the differences between the theoretical direct gaps and the optical gap are used,
since no value of the experimental fundamental gap for FeO is available.
The Hubbard parameters $U$ and $J$ values used  for $G_0W_0$@LDA+$U$
and $GW_0$@LDA+$U$ were obtained from constrained DFT calculations~\cite{PhysRevB.82.045108}.
$M$ denotes metal.
}
\begin{ruledtabular}
\begin{tabular}{lccccccccc}
         &   \multicolumn{2}{c}{MnO}  &   \multicolumn{2}{c}{FeO}       &  \multicolumn{2}{c}{CoO}   &   \multicolumn{2}{c}{NiO}  & MAPE \\
       \cline{2-3}  \cline{4-5} \cline{6-7}  \cline{8-9}
         &   $E^i_{\rm g}$ & $E^d_{\rm g}$  &   $E^i_{\rm g}$ & $E^d_{\rm g}$  &   $E^i_{\rm g}$ & $E^d_{\rm g}$  &   $E^i_{\rm g}$ & $E^d_{\rm g}$ \\
         \hline
This work \\
 PBE      &    0.84   & 1.38         & $M$  & $M$  & $M$ &  $M$ &  0.97  & 1.13  & --- \\
 PBE0    &   3.65    &  4.29        &  3.02    & 3.42            &  4.25    & 5.05    &  5.29  & 5.82    & 37\% \\
 HSE03  &   2.65    & 3.28        &   2.11    & 2.68            &  3.21    & 3.96    &   4.28  & 4.74   & 19\% \\
 HSE06  &   2.92    & 3.56        &   2.27    & 2.66            &  3.50    & 4.29    &   4.56  & 5.06   & 22\% \\
 MHSE   &   4.09    & 4.70        &   3.54    & 3.94            &  4.90    & 4.98    &   5.72  & 6.38   & 50\% \\
 DD0-RSH-CAM  &   3.61     & 4.23       &   2.27    & 2.63              &  3.01    & 4.14    &  4.34  & 4.99  & 11\% \\
 DSH0                    &   3.37     & 3.99       &   2.11    & 2.40              & 2.90     & 4.02    &  4.16   & 4.82 & ~~8\% \\
 DD-RSH-CAM    &   4.93      & 5.57      &  5.09      & 5.40             &  5.61     & 6.70    &   6.34  & 7.02 & 81\% \\
 DSH                      &   4.43      & 5.07      &   4.73    & 5.05              &  5.23     & 6.51    &   5.91 & 6.59  & 68\% \\
 $G_0W_0$@PBE   & 1.60  & 2.07 & $M$  & $M$  & $M$ &  $M$ & 1.58 & 1.76 & --- \\
 ev$GW_0$@PBE & 1.92  & 2.38 & $M$  & $M$  & $M$ &  $M$ & 1.78 & 2.18 & --- \\
 \\
 Other theoretical work \\
$G_0W_0$@HSE03~\cite{PhysRevB.79.235114} & 3.4  & 4.0 &2.2 & 2.3 & 3.4 & 4.5 & 4.7 & 5.2 & 18\% \\
$G_0W_0$@LDA+$U$~\cite{PhysRevB.82.045108} &2.34 &  & 0.95 &  & 2.47 & & 3.75  &  & --- \\
ev$GW_0$@LDA+$U$~\cite{PhysRevB.82.045108} & 2.57  &  & 0.86 & & 2.54 & & 3.76 &  & --- \\
\\
Experiment \\
PES+BIS  &  3.9$\pm$0.4\cite{PhysRevB.44.1530}&  &   &    &2.5$\pm$0.3\cite{PhysRevB.44.6090} &   & 4.3\cite{PhysRevLett.53.2339}  &  &  \\
XAS+XES &  4.1\cite{PhysRevB.77.165127}   &  &    &     &    2.6~\cite{PhysRevB.77.165127} &  &   4.0\cite{PhysRevB.77.165127}  & &  \\
Optical absorption   &    & 3.6$\sim$3.8\cite{MnO1968} &    & 2.4\cite{BOWEN1975355}   &    & 2.7~\cite{PhysRevB.2.2182} &    &3.7\cite{PhysRevB.2.2182} &   \\
  &    &  &    &    &    & 5.43\cite{jkps.50.632} &    & 3.87\cite{jkps.50.632} &  \\
\end{tabular}
\end{ruledtabular}
\label{Table: band_gap_compare}
\end{table*}

Table~\ref{Table: beta_mu} compiles the parameters $\beta$ and $\mu$
used in single-shot DD0-RSH-CAM and DSH0, and self-consistent DD-RSH-CAM and DSH calculations.
It can be seen that DD0-RSH-CAM and DSH0 have very similar parameters $\mu$, though $\mu$ are
computed in different ways. This is further manifested in Fig.~\ref{fig:eps_G_all}, where the model
dielectric functions of DD0-RSH-CAM and DSH0 almost match for FeO, CoO and NiO.
In addition, the RPA@PBE screenings of all considered materials are well captured by DD0-RSH-CAM and DSH0.

Table~\ref{Table: band_gap_compare} shows the resulting calculated indirect and direct band gaps from
different methods,  along with the experimental gaps obtained from
X-ray photoemission spectroscopy (XPS) and Bremsstrahlung isochromat spectroscopy (BIS)
~\cite{PhysRevB.44.1530,PhysRevB.44.6090,PhysRevLett.53.2339},
oxygen $K\alpha$ X-ray emission spectroscopy (XES) and oxygen 1$s$ X-ray absorption spectroscopy (XAS)~\cite{PhysRevB.77.165127},
as well as  optical absorption~\cite{MnO1968,BOWEN1975355,PhysRevB.2.2182,PhysRevB.2.2182,jkps.50.632}.
Since an accurate determination of band gaps in experiment is not trivial and accuracy is affected by complicated experimental factors,
such as sample qualities, instrumental resolutions, mixtures of bulk and surface, excitonic effects in optical absorption, electron-phonon coupling, and so on,
a direct comparison between the theoretical and experimental gaps should be done cautiously.
However, from Table~\ref{Table: band_gap_compare} one can see that the fundamental gaps obtained from XAS+XES  are quite close to those from XPS+BIS
and therefore we compare the theoretical indirect gaps to the experimental fundamental gaps for the assessment of different methods.
For FeO where no experimental fundamental gap is available, the optical gap from the optical absorption~\cite{BOWEN1975355}
is used to compare to the theoretical direct gaps. Certainly, excitonic effects will reduce the direct gaps,
but to consider such effects, one needs to solve the BSE, which goes beyond the scope of this work.

As shown in Table~\ref{Table: band_gap_compare}, due to the strong SIE for localized electrons, PBE underestimates the band
gaps for MnO and NiO and even wrongly gives metallic solutions for FeO and CoO.
The screened HSE03 functional opens the band gaps for FeO and CoO owing to the
inclusion of one quarter of the SR Fock exchange, reducing the SIE.  The predicted gap for NiO is about 4.28 eV,
which is in excellent agreement with the experimental gap (4.3 eV~\cite{PhysRevLett.53.2339}).
Nevertheless, HSE03 underestimates the gap of MnO by 1.2 eV. Our HSE03 results
are consistent with published data~\cite{0953-8984-20-6-064201, PhysRevB.79.235114}.
Decreasing the range-separation parameter $\mu$ from $0.3~\AA^{-1}$ (HSE03) to $0.2~\AA^{-1}$ (HSE06)
increases the gaps for all compounds, improving the gaps for MnO and FeO towards experiments but deteriorating
the agreement with experiments for CoO and NiO. Moreover,  the gap for MnO is still too small.
It is known that increasing the SR Fock exchange fraction $\alpha$ increases the gap, while increasing $\mu$ decreases
the gap~\cite{Franchini_2014}. One would thus expect that a suitable tuning of $\alpha$ and $\mu$ parameters might yield improved band gaps.
MHSE with $\alpha=0.6$ and $\mu=0.5~\AA^{-1}$ suggested in Ref.~\cite{RevModPhys.86.253}
now yields a very good band gap for MnO compared to experiment. Nevertheless, it increases the gaps
for the other three compounds significantly, making the agrement with experiments much worse.
This also implies that the effect of increased $\alpha$ on the gap wins over the one induced by increasing $\mu$ in MHSE.
Compared to MHSE, PBE0, which includes one quarter of the full-range Fock exchange, gives relatively smaller gaps,
but the predicted band gaps for FeO, CoO and NiO are still too large compared to experiments.

The fact that all above-mentioned conventional hybrid functionals fail to accurately describe the band gaps
for all compounds simultaneously arises from the inadequate description of the momentum-dependent dielectric function (see Fig.~\ref{fig:HSE_eps_G}).
Single-shot DD0-RSH-CAM and DSH0, on the other hand, well reproduce the RPA screening from first-principles
for the entire momentum range (see  Fig.~\ref{fig:eps_G_all}).
This makes DD0-RSH-CAM and DSH0 superior to conventional hybrid functionals in the overall description of band gaps.
As shown in Table~\ref{Table: band_gap_compare}, compared to HSE03 and HSE06,
DD0-RSH-CAM and DSH0 improve the gaps of MnO and CoO towards experiments but without
destroying the results for FeO and NiO, leading to the smallest MAPE among the considered functionals.
Compared to DSH0, the larger gaps predicted by DD0-RSH-CAM are due to the smaller $\mu$  (see Table~\ref{Table: beta_mu}).
The good performance of DSH0 as DD0-RSH-CAM also implies that
Eq.~(\ref{eq:mu_for_DSH}) is indeed a good approximation to determine the parameter $\mu$.

\begin{figure}[h!]
\begin{center}
\includegraphics[width=0.49\textwidth,clip]{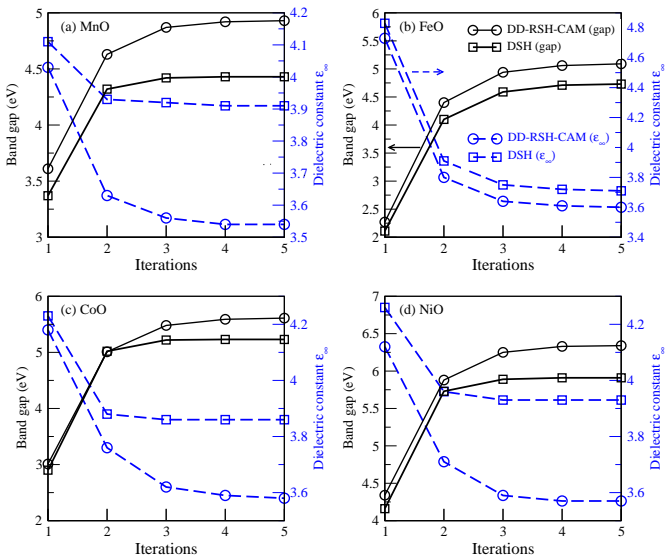}
\end{center}
 \caption{(Color online)  Evolution of the dielectric constant $\epsilon_\infty$ (dashed lines and right-hand axis)
 and band gap (in eV) (solid lines and left-hand axis)
as a function of the number of iterations. For FeO, the finite field approach to calculate $\epsilon_\infty$ for DSH
failed to converge, hence we applied to same procedure as for DD-RSH-CAM to determine $\epsilon_\infty$, which
explains why  DD-RSH-CAM  and DSH are so close for FeO.
}
\label{fig:scf}
\end{figure}

In contrast to what is observed in Refs.~\cite{PhysRevMaterials.2.073803, doi:10.1021/acs.jpclett.8b00919}-- namely,
that self-consistent DD-RSH-CAM or DSH perform almost equally well as the single-shot counterparts
for the considered data sets --here we find that self-consistency deteriorates the results
with a significant overestimation of band gaps due to underestimated dielectric constants (see Fig.~\ref{fig:scf}).
Moreover, it was found in Ref.~\cite{PhysRevMaterials.2.073803} that  self-consistent DD-RSH-CAM delivers
slightly better numbers compared to DD0-RSH-CAM (MAPE=7.3\% vs. MAPE=7.9\%, respectively),
which is apparently in contrast with the  results obtained in Ref.~\cite{doi:10.1021/acs.jpclett.8b00919}
where self-consistent DSH was found to yield a MAPE nearly twice as large as DSH0 (9\% against 5\%, respectively).
However, a closer inspection of the data in Refs.~\cite{PhysRevMaterials.2.073803, doi:10.1021/acs.jpclett.8b00919}
shows that oxide compounds such as Cu$_2$O, In$_2$O$_3$, ZnO, TiO$_2$, and MgO
show larger errors compared to the other members of the considered data set
for self-consistent DD-RSH-CAM and DSH~\cite{PhysRevMaterials.2.073803, doi:10.1021/acs.jpclett.8b00919},
suggesting that self-consistent DD-RSH-CAM and DSH might not be suitable for
the prescription of the band gaps of oxides. This needs to be verified by more data on oxides.
We note that including a bootstrap $f^{\rm boot}_{\rm xc}$ in the full response function
increases the dielectric constants compared to RPA,
but the calculated dielectric constants are still smaller than those predicted by the finite field approach
(compare DD-RSH-CAM to DSH for MnO, CoO, and NiO in Fig.~\ref{fig:scf}).
In addition,  DD-RSH-CAM delivers smaller $\mu$ parameters than DSH
(see Table~\ref{Table: beta_mu}). All the enumerated points explain
the larger band gaps predicted by DD-RSH-CAM compared to DSH.

It is remarkable that single-shot $G_0W_0$@PBE and eigenvalue self-consistent ev$GW_0$@PBE
give much too small band gaps for MnO and NiO (see Table~\ref{Table: band_gap_compare}).
The reason for this is that using PBE, the dielectric constants are
overestimated as shown in Table~\ref{Table: dielectric_compare}, and too large screening ($W$) usually yields too small
band gaps in the $GW$ approximation. Using HSE as starting point improves the situation~\cite{PhysRevB.79.235114}, since
HSE yields  larger one-electron band gaps to start with, as well as a significantly reduced screening,
both resulting in larger final $GW$ band gaps. Likewise, using LDA+$U$ as starting point also increases
the initial one-electron band gap  and reduces the screening. However,
the calculated gaps for $G_0W_0$@LDA+$U$ and ev$GW_0$@LDA+$U$ for these four
compounds exhibit strong dependence on the Hubbard parameters $U$ and $J$ used in the LDA+$U$ calculations~\cite{PhysRevB.82.045108}.
Using the Hubbard parameters computed from constrained DFT calculations for $G_0W_0$@LDA+$U$
and $GW_0$@LDA+$U$ yields too small gaps for FeO (see Table~\ref{Table: band_gap_compare}).

The most interesting comparison is between ev$GW_0$ and single-shot DD0-RSH-CAM/DSH0, since
all three apply a very similar momentum-dependent non-local exchange.
The single-shot DD0-RSH-CAM/DSH0 yields remarkably accurate band gaps, whereas ev$GW_0$
clearly underestimates the band gap.
To exclude that  orbital relaxation (present in the hybrid calculations) is the source of the difference, we also performed DD0-RSH-CAM calculations using fixed PBE orbitals
and applying first order perturbation theory. This yields within 200 meV the same gaps as in Table~\ref{Table: band_gap_compare}.
Thus, orbital relaxation is not responsible for the difference between DD0-RSH-CAM/DSH0 and ev$GW_0$@PBE.
Hence, the difference is a genuine effect of the different treatment of correlation effects.
The hybrid functionals obviously completely neglect dynamic correlation effects (frequency dependency) but
include terms beyond the Hartree screening (i.e., RPA diagrams). We believe that the good agreement of DD0-RSH-CAM/DSH0
with experiment is  to some extent fortuitous, since the $\epsilon_\infty$ provided by PBE
(compare Table~\ref{Table: dielectric_compare}) is way too large. This also means
that  DD0-RSH-CAM/DSH0 uses too little exact exchange, and should concomitantly underestimate
the gap as does ev$GW_0$@PBE. Nevertheless,  it is convenient
that such a simple approximation works so remarkably well.

\begin{table}
\caption {Ion-clamped macroscopic dielectric constants calculated
from different functionals as well as available experimental values.
For PBE calculations, density functional perturbation theory
(DFPT)~\cite{RevModPhys.73.515,PhysRevB.73.045112} is used,
whereas for the hybrid functionals, the finite field approach
is employed. Both include the effects of exchange and correlation exactly.
For FeO, where the finite field approach failed to converge, the Dyson equation was solved
including the  RPA and local field effects via the bootstrap $f_{\rm xc}$ kernel.
}
\begin{ruledtabular}
\begin{tabular}{lcccc}
         &   MnO  &   FeO       &  CoO   &   NiO  \\
         \hline
 PBE      &     7.96    & ---   & --- &  17.22  \\
 PBE0    &   4.52            &  4.51              &  4.83      &   5.03  \\
 HSE03  &   4.59            &   5.18                &  4.91       &   5.13 \\
 HSE06  &   4.54          &   5.10               &  4.84       &   5.05 \\
 MHSE   &   3.71           &   3.96                &  4.71       &   3.74  \\
 DD0-RSH-CAM  &   4.03           &   4.73                 &  4.18      &  4.12  \\
 DSH0     &   4.11          &   4.83                & 4.23       &   4.26  \\
 DD-RSH-CAM  &   3.54          &  3.60          &  3.58   &   3.57 \\
 DSH     &   3.91           &   3.71          &  3.86   &   3.93 \\
 Expt.   & 4.95\cite{PLENDL1969109}  & ---  & 5.3\cite{doi:10.1063/1.1714508} & 5.7\cite{doi:10.1063/1.1714508}
\end{tabular}
\end{ruledtabular}
\label{Table: dielectric_compare}
\end{table}

\begin{table}
\caption {Magnetic moments (in $\mu$B/atom) from different functionals as well as available experimental values. }
\begin{ruledtabular}
\begin{tabular}{lcccc}
         &   MnO  &   FeO       &  CoO   &   NiO  \\
         \hline
 PBE      &     4.16    & 3.32   & 2.40 &  1.37  \\
 PBE0    &   4.36             &  3.50               &  2.64      &   1.69  \\
 HSE03  &   4.36            &   3.49               &  2.63       &   1.68 \\
 HSE06  &   4.36          &   3.50               &  2.64       &   1.68  \\
 MHSE   &   4.47           &   3.60                &  2.73       &   1.80  \\
 DD0-RSH-CAM  &   4.46           &   3.56                &  2.70      &   1.79  \\
 DSH0     &   4.45          &   3.55                & 2.70       &   1.78  \\
 DD-RSH-CAM  &   4.48          &  3.59          &  2.73   &   1.81 \\
 DSH     &   4.46          &   3.58          &  2.73   &   1.80 \\
 Expt.   & 4.58\cite{PhysRevB.27.6964}  & 3.32\cite{PhysRev.110.1333} & 3.35\cite{PhysRevB.1.2243} & 1.64\cite{Alperin1962} \\
    &    & 4.20\cite{0022-3719-12-2-021}  & 3.98\cite{PhysRevB.64.052102} & 1.90\cite{PhysRevB.27.6964,PhysRev.110.1333}
\end{tabular}
\end{ruledtabular}
\label{Table: moments_compare}
\end{table}

We are now turning to assess how the DD-RSH-CAM and DSH hybrid functionals perform on
other properties such as dielectric constants, magnetic moments, and DOS.
Table~\ref{Table: dielectric_compare} reports the calculated ion-clamped macroscopic
dielectric constants $\epsilon_\infty$ from different functionals.
As expected, the larger the band gap is, the smaller is $\epsilon_\infty$ (compare Table~\ref{Table: band_gap_compare} with Table~\ref{Table: dielectric_compare}).
Although DD0-RSH-CAM and DSH0 yield the best band gaps, their predicted $\epsilon_\infty$ are less satisfactory (too small) than those
predicted by PBE0, HSE03, and HSE06 compared to experiments. Self-consistent DD-RSH-CAM and DSH give an even worse
description of $\epsilon_\infty$ for the considered compounds due to the significant overestimation of the band gaps.

\begin{figure}
\begin{center}
\includegraphics[width=0.44\textwidth,clip]{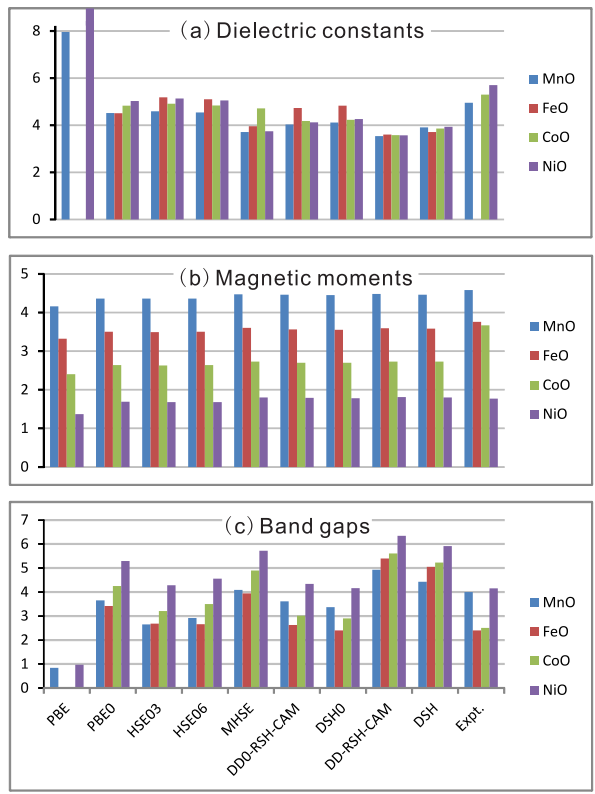}
\end{center}
 \caption{(Color online) Comparison of (a) dielectric constants, (b) magnetic moments (in $\mu$B/atom),
 and (c) fundamental band gaps (except for FeO) (in eV) calculated from different functionals with respect to experimental
 values. Again, due to the unavailability of the experimental fundamental gap of FeO, in (c) for FeO we have shown the comparison
 of the theoretical direct gaps with its experimental optical gap~\cite{BOWEN1975355}.
}
\label{fig:histogram}
\end{figure}

Table~\ref{Table: moments_compare} compiles the calculated magnetic moments from different functionals
as well as experimental values. As expected,  PBE underestimates the magnetic moments due to  overdelocalization of electrons.
Hybrid functionals increase the magnetic moments
by about 5$\sim$10\% as compared to PBE,  improving the agreement with experiments.
Also, one can see that magnetic moments are not so sensitive to the adopted hybrid functional.
It is worth noting that the orbital contribution to the magnetic moment is not included,
since the spin-orbit coupling is not considered in our calculations.
Therefore, the larger deviation between the theoretical and experimental magnetic
moments for CoO (Table~\ref{Table: moments_compare}) is
ascribed to the missing large orbital moment ($\sim$ 1 $\mu$B)~\cite{doi:10.1143/JPSJ.67.2505,RADWANSKI2004107}.
For the sake of better comparison,
Fig.~\ref{fig:histogram} histogramatically displays the calculated magnetic moments, dielectric constants,
and band gaps for all different functionals against experimental values for the four compounds.

\begin{figure}
\begin{center}
\includegraphics[width=0.492\textwidth,clip]{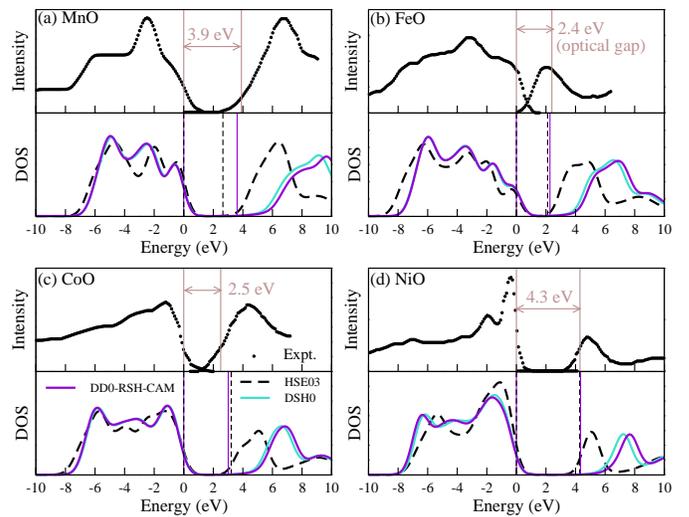}
\end{center}
 \caption{(Color online) Calculated DOS (bottom panels) of (a) MnO, (b) FeO, (c) CoO, and (d) NiO from HSE03, DSH0, and DD0-RSH-CAM
  hybrid functionals compared to experimental XPS and BIS spectra
  (upper panels)~\cite{PhysRevB.44.1530, PhysRevB.44.6090, PhysRevLett.53.2339,0953-8984-11-7-002}.
  The top of the theoretical valence bands has been set to energy zero and
  the experimental spectra have been aligned to the theoretical DOS in terms of the upper valence band edge.
  The vertical lines indicate the experimental band gaps (upper panels) and theoretical band gaps
  (bottom panels, black dashed lines for HSE03 and solid violet lines for DD0-RSH-CAM).
For better comparison with experimental spectra, the calculated DOS are broadened by a Gaussian with 0.5 eV full width at half maximum.
Note that since Zimmermann \emph{et al.}~\cite{0953-8984-11-7-002} did not derive a value
for the gap of FeO from their XPS+BIS spectra,  the gap from optical absorption~\cite{BOWEN1975355} is shown in (b).
}
\label{fig:DOS}
\end{figure}

Fig.~\ref{fig:DOS} shows the calculated DOS from HSE03, DSH0, and DD0-RSH-CAM
hybrid functionals compared to experimental XPS+BIS spectra. One can see that
DSH0 predicts very similar DOS as DD0-RSH-CAM due to their very similar $\mu$ parameters and
model dielectric functions used. The valence band DOS predicted for DSH0 and DD0-RSH-CAM
are as satisfactory as HSE03, which are in good agreement with the experimental
spectra~\cite{PhysRevB.44.1530, PhysRevB.44.6090, PhysRevLett.53.2339,0953-8984-11-7-002}.
However, the main peaks in the conduction band (CB) DOS are shifted to higher energies by about 2 eV by DSH0 and DD0-RSH-CAM compared to HSE03,
though their calculated band gaps for FeO, CoO, and NiO are similar.
Our obtained DOS from DSH0 and DD0-RSH-CAM are consistent with those calculated from $G_0W_0$@HSE03~\cite{PhysRevB.79.235114},
which also predicts too large shifts for the CB DOS.
The HSE03 calculated CB DOS seems to be in better agreement with experimental BIS spectra compared to DSH0 and DD0-RSH-CAM.
In particular for NiO, the main peak in the BIS spectra at around 5 eV is
well reproduced by HSE03,  whereas DSH0 and DD0-RSH-CAM
shift upward this peak by about 2 eV,  deviating from the BIS spectra [Fig.~\ref{fig:DOS}(d)].
However, caution needs to be taken when comparing theory and experiment for FeO and CoO.
For instance, the experimental XPS and BIS spectra for CoO exhibit
large broadening (1.0 eV for the XPS and 0.8 eV for the BIS). This yields
a photoemission gap of 2.5$\pm$0.3 eV~\cite{PhysRevB.44.6090} that is smaller than our
calculated band gaps by DSH0 and DD0-RSH-CAM [see Fig.~\ref{fig:DOS}(c)].
Furthermore, for FeO, the XPS and BIS spectra~\cite{0953-8984-11-7-002} do not seem to be properly aligned,
since the estimated band gap from the XPS+BIS spectra is much lower than even  the optical gap of 2.4 eV~\cite{BOWEN1975355}.
The experimental optical gap is in reasonable agreement with our calculated band gaps for DSH0 and DD0-RSH-CAM [see Fig.~\ref{fig:DOS}(b)].

\section{Conclusions}\label{sec:conlcusions}

In conclusion, we have discussed the connection between different hybrid functionals
using the gKS formalism, and we have assessed the performance of the recently proposed DD-RSH-CAM and DSH functionals
in their single-shot and self-consistent versions on challenging
antiferromagnetic transition-metal monoxides (TMOs) MnO, FeO, CoO, and NiO. We have evaluated
the band gaps, the electronic density of states, the dielectric constants, and magnetic moments
by comparing them to other conventional hybrid functionals such as PBE0, HSE03, and HSE06,
as well as a modified HSE with $\alpha=0.6$ and $\mu=0.5~{\AA}^{-1}$ and the $GW$ method.
We have emphasized that the DD-RSH-CAM and DSH hybrid functionals have essentially the same functional form
for the exchange-correlation potential with a common model dielectric function.
DSH is parameterized by determining the exact long-range dielectric constant within the applied density
functional theory approximation. DD-RSH-CAM is somewhat more approximate and uses either
RPA screening or the so-called bootstrap kernel for the exchange and correlation effects to determine the dielectric screening.  The range separation parameter
$\mu$ is also obtained in a different manner for both hybrid functionals, where
DD-RSH-CAM is slightly more rigorous and fits the parameter to the wave vector dependent screening. Despite these differences,
we find that single-shot DD0-RSH-CAM and DSH0 perform almost equally
and both hybrid functionals outperform conventional hybrid functionals
for the band gaps with the smallest MAPE compared to experimental gaps.
This is in line with the finding reported in Refs.~\cite{PhysRevMaterials.2.073803,doi:10.1021/acs.jpclett.8b00919}
based on a large dataset of conventional insulators and semiconductors.
 In fact, for the TMOs series DD0-RSH-CAM and DSH0 yield band gaps in excellent agreement with experiment
 and are capable to qualitatively predict the band gaps, a feat that
is not even achieved by the $G_0W_0$@PBE  and ev$GW_0$@PBE approximation.
The key achievement of
the two new functionals is that they model the momentum-dependent screened exchange accurately and complement this
with a semi-local functional that restores the well known constraints for the exchange and
correlation hole. We note that
the static screened exchange plus Coulomb hole method is similar in spirit and arguably
more general insofar that the entire non-local screening tensor is accounted for. However,
the description of the static Coulomb hole (which is a local potential) is firmly rooted in the random phase approximation,
which observes less sum rules and known constraints than density functionals.

Let us now turn to those aspects that are less satisfactory. First, self-consistent DD-RSH-CAM and DSH
 deteriorate the results with a significant overestimation of the band gaps. This issue was
not observed in previous studies and might well be related to the correlated character of
TMOs. Second and to some extent related,  it is found that DD0-RSH-CAM and DSH0 (even more so DD-RSH-CAM and DSH)
underestimate the dielectric constants compared to experiment.  This leads to  the overestimation of the
gaps in the self-consistent description, since a too small dielectric constant implies too much exchange
and thus a too large band gap. Clearly, this aspect needs further considerations.
Finally, although the valence band DOS predicted by
DD0-RSH-CAM and DSH0 are as satisfactory as for, e.g., HSE03 in reproducing the experimental spectra,
the predicted conduction band DOS for DD0-RSH-CAM and DSH0
are shifted to higher energies by about 2 eV compared to those predicted by HSE03. Whether this
worsens or improves agreement with experiment is a matter of debate: it is very
difficult to align the conduction band and valence band spectra, as they are determined independently experimentally.
Hence any conclusions must be drawn rather carefully. Overall, however the conduction band spectra
of DD0-RSH-CAM and DSH0  seem to be slightly blue shifted compared to experiment. This  could be
a result of the neglect of dynamic correlation effects (frequency dependency of the self-energy reduces the
band width)
or related to the simple diagonal approximation used in the screening.

In summary, considering the very good performance of the nonempirical DD0-RSH-CAM and DSH0 functionals
in predicting band gaps for various narrow-, intermediate-, and wide-gap semiconductors
and insulators~\cite{PhysRevMaterials.2.073803,doi:10.1021/acs.jpclett.8b00919}
 as well as challenging transition-metal monoxides, DD0-RSH-CAM and DSH0 are undoubtedly
promising hybrid functionals for band-gap related applications such as band alignments
and optical properties. On the other hand, the materials dependent parameters
make these functionals difficult to apply to  the description of, e.g.,
hetero-structures, combinations of different materials characterized by very different screening, surfaces, as well as
extended defects where the screening changes substantially compared to bulk phases. We also feel that the
good performance of DD0-RSH-CAM/DSH0 fitted to PBE screening is somewhat fortuitous, since PBE substantially
overestimates the long-range screening in all transition-metal oxides, which explains the too small
$G_0W_0$@PBE band gaps predicted for TMOs. So it is a lucky coincidence that DD0-RSH-CAM and DSH0 work so well
across the board: DD0-RSH-CAM/DSH0 are an excellent pragmatic solution but theoretically somewhat
unsatisfactory.

\section*{Acknowledgements}

P. Liu thanks H. Jiang for sharing his scripts on DSH hybrid functional and
thanks S. Sharma for helpful discussions on the bootstrap kernel.
This work was supported by the Austrian Science Fund (FWF) within the SFB ViCoM (Grant No. F 41).
Supercomputing time on the Vienna Scientific cluster (VSC) is gratefully acknowledged.

\bibliographystyle{apsrev4-1}
\bibliography{reference} 

\end{document}